\documentstyle[aps,prl,epsfig]{revtex}
\twocolumn
\newcommand{\f}{\frac}
\newcommand{\be}{\begin{equation}}
\newcommand{\ee}{\end{equation}}
\newcommand{\timesring}{{\otimes}}

\title{Loop algorithms for asymmetric Hamiltonians}
\author{Olav F. Sylju{\aa}sen\footnote{email: sylju@nordita.dk}}
\address{Massachusetts Institute of Technology, 77 Massachusetts Avenue,
         Cambridge, MA 02139, USA \\
         NORDITA, Blegdamsvej 17, DK-2100 Copenhagen {\O},
         Denmark}

\begin{document}
\maketitle

\begin{abstract}
Generalized rules for building and flipping clusters in 
the quantum Monte Carlo loop algorithm are presented for the 
XXZ-model in a uniform magnetic field along the Z-axis. 
As is demonstrated for the Heisenberg antiferromagnet it is possible from these rules to select a
new algorithm which performs significantly better than the standard loop algorithm in strong 
magnetic fields at low temperatures.
\end{abstract}

\section{Introduction}
The invention of the Loop algorithm\cite{Evertz} was a major breakthrough for Monte Carlo
simulations of quantum spin systems. This algorithm which is a quantum version of the
Swendsen-Wang cluster algorithm\cite{Kawashima} has many desirable features which makes it possible to study large
systems at low temperatures\cite{Depleted}. 
Among them is that the algorithm can be formulated directly in continuous imaginary time\cite{Beard}, thus
avoiding the need to extrapolate data obtained at finite imaginary time discretization.  
Another is that updates are done in an extended configuration space of spins and new entities called loops.
This makes big changes in the spin configuration possible in a single Monte Carlo step, resulting in very small
autocorrelation times. Furthermore the nonlocal updating procedure allows all topological sectors to be sampled. For a quantum spin system this means in particular that sectors with different magnetization are sampled. This is in contrast to
most other algorithms which operate at fixed magnetization. 

Although excellent for a wide class of models, 
the Loop algorithm does not do well when the Hamiltonian is made asymmetric by a uniform magnetic field or a chemical potential. In these cases autocorrelation times become very long at low temperatures and the performance of the algorithm is lowered drastically. Here it is shown how this can be overcome
by generalizing the loop algorithm. The
generalization is obtained by relaxing the condition of non-interacting loops, and
by taking the magnetic field into account in the loop building process.
To be specific, we consider the nearest neighbor XXZ-model 
on a bipartite lattice in a magnetic field along the Z-axis
\be
 {\cal H} = \sum_{<ij>} \left( J_x S^x_i S^x_j +J_x S^y_i S^y_j + J_z S^z_i S^z_j                    \right)-H \sum_i S^z_i.
\ee
Despite this choice of model it is expected that the procedure employed here should apply to other quantum models as well, such as lattice fermions in the presence of a chemical potential.
In the next section the generalized loop algorithm is presented. Then it is explained how an algorithm that
performs well in a magnetic field can be chosen. The usefulness of this algorithm  
is demonstrated by measuring magnetization curves for the Heisenberg antiferromagnet on a dimer, a chain, and on a plane. Finally it is shown that the algorithm can also be used to determine
the critical temperature for the Kosterlitz-Thouless transition occuring at finite magnetic fields in
the Heisenberg antiferromagnet.

\section{The Algorithm}

To explain the algorithm we begin by formulating the $d$ dimensional quantum system as a classical system in $d+1$ dimensions. This is done in the
standard way\cite{Evertz} of dividing the Hamiltonian into sums of commuting pieces. Then a Trotter-Suzuki breakup is performed, and complete
sets of states, which are labelled by their eigenvalues for $S^z_i$, are inserted at each time-slice between each
sum of commuting pieces. 
The matrix elements are easily evaluated
and corresponds to interactions around shaded plaquettes in a generalized checkerboard pattern. 

\begin{figure}
\begin{center}
\epsfig{file=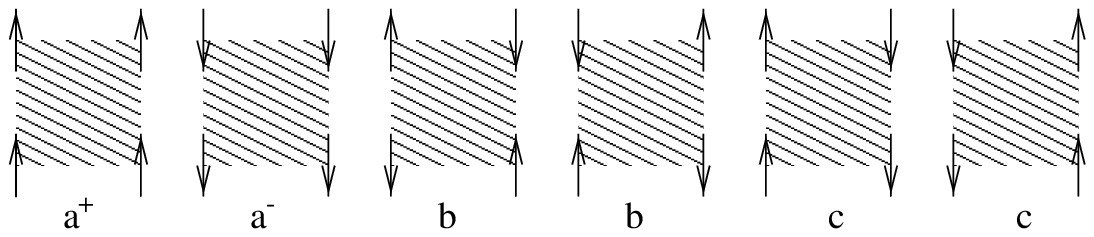,width=7.0cm}
\end{center}
\caption{The different plaquettes for the XXZ-model in a magnetic field. The vertical is the imaginary time direction. \label{weights}}
\end{figure}

As is shown in fig. \ref{weights} 
there are six allowed spin configurations around a shaded plaquette
for the XXZ-model in a magnetic field. Other configurations have zero weights as for those $S^z$ is not conserved along the time direction. 

Because the loop algorithm can be formulated in continuous imaginary time
it is sufficient to consider the limit where the imaginary time spacing $\Delta \tau$ goes to zero. In this limit the plaquette weights are  
\begin{eqnarray}
     w(a_+) & = & 1- \left( \f{J_z}{4}-\f{H}{z} \right) \Delta \tau, \nonumber \\
     w(a_-) & = & 1- \left( \f{J_z}{4}+\f{H}{z} \right) \Delta \tau, \\
     w(b)   & = & \f{|J_x|}{2} \Delta \tau, \nonumber \\
     w(c) & = & 1+ \f{J_z}{4} \Delta \tau, \nonumber
\end{eqnarray}
where $z$ is the lattice coordination number. 

\begin{figure}
\begin{center}
\epsfig{file=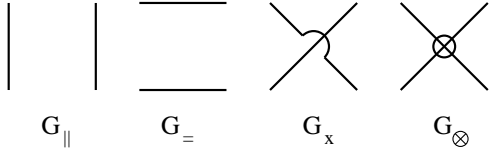,width=6.5cm}
\end{center}
\caption{The different breakups $G$ into loop segments around a shaded plaquette.\label{segments}}
\end{figure}

The loop algorithm consists of two main steps.
The first is to build
loops. Loop building is a probabilistic process where each shaded plaquette $p$ is broken up into loop segments $G_p$ with a probability $P(s_p \rightarrow s_p,G_p)$, dependent
on the spin configuration $s_p$. 
Each loop segment connects two or four spins.
The different types of loop segments are shown in fig.~\ref{segments}.
When this is done for all shaded plaquettes, the entire space-time lattice will be filled with loops. The second step is to flip spins along one or more loops. Because of the way the break-ups are constructed, this always results in an allowed spin configuration provided all the spins around a loop are flipped.  The process of flipping the spins along a loop is also probabilistic.
It is governed by the probability $P_G(s \rightarrow s\prime)$ for changing spin configurations given a particular break-up $G$ for the whole lattice.
After this second step a new spin configuration is generated and one does
the measurements and start over again.
For the whole procedure to satisfy detailed balance it is sufficient\cite{Kandel} that
the probabilities are chosen such that
\begin{eqnarray}
   P(s_p \rightarrow s_p,G_p) & = & \f{w(s_p,G_p)}{w(s_p)}, \label{rule1} \\
   P_G(s \rightarrow s\prime) \prod_p w(s_p,G_p) & = & P_G(s\prime \rightarrow s )
   \prod_p w(s_p\prime,G_p), \label{rule2}
\end{eqnarray}
where $G$ and $s$ are the full loop and spin configuration, pieced together by
the loop segments $G_p$ and plaquette spins $s_p$ respectively. 
$w(s_p,G_p)$ is the plaquette weight of plaquette $p$ in the extended configuration space of
both spins and loops. The weights $w(s_p,G_p)$ must be positive definite and satisfy
\be \label{rule3}
   \sum_{G_p} w(s_p,G_p) = w(s_p).
\ee
The different loop algorithms described here correspond to different choices of these weights.
Writing out Eq.~(\ref{rule3}) explicitly we find
\begin{eqnarray}
   w(a_+) & = & w(a_+,G_{||}) + w(a_+,G_\times)+w(a_+,G_\timesring) ,\label{eq6} \\
   w(a_-) & = & w(a_-,G_{||}) + w(a_-,G_\times)+w(a_-,G_\timesring), \\
   w(b) & = & w(b,G_{=}) + w(b,G_\times) + w(b,G_\timesring), \\
   w(c) & = & w(c,G_{||}) + w(c,G_{=}) + w(c,G_\timesring). \label{eq9}
\end{eqnarray}
We have set the weights $w(a_+,G_{=})$, $w(a_-,G_{=})$, $w(b,G_{||})$
and $w(c,G_\times)$ to zero as flipping the spins along one loop segment
for such configurations leads to a configuration with zero weight. 
It is therefore clear that we have eight parameters at our disposal. Let
us parametrize the weights in the following way
\begin{eqnarray}
   w(a_+,G_\times) & = & s \Delta \tau, \\
   w(a_-,G_\times) & = & t \Delta \tau, \\
   w(b,G_=) & = & u \Delta \tau, \\
   w(c,G_=) & = & v \Delta \tau, \\
   w(a_+,G_\timesring) & = & e \Delta \tau, \\
   w(a_-,G_\timesring) & = & f \Delta \tau, \\
   w(b,G_\timesring) & = & g \Delta \tau, \\
   w(c,G_\timesring) & = & h \Delta \tau. 
\end{eqnarray}
The remaining four weights are given by Eqs.~(\ref{eq6})-(\ref{eq9}).
Note that although this is a convenient way of 
parametrizing the weights it is not the most general one. 
In selecting the parametrization above we have chosen
which weights are of order $\Delta \tau$ or 1. 

With this parametrization it is easy to obtain the loop building probabilities $P(s_p \rightarrow s_p,G_p)$
from Eqs.~(\ref{rule1}) and (\ref{rule3}).
To have non-negative weights, all the parameters must be greater than or equal to zero and $(u+g) \le |J_x|/2$.

To satisfy detailed balance in the extended con\-figuration space 
Eq.~(\ref{rule2}) we need the ratios $w(s_p\prime,G_p)/w(s_p,G_p)$
which are
\begin{eqnarray}
   \f{w(c,G_{||})}{w(a_+,G_{||})} & = & 1+\Delta \tau (
     \f{J_z}{2}-\f{H}{z} + s+e -v-h), \label{first-flip} \\
   \f{w(c,G_{||})}{w(a_-,G_{||})} & = & 1+\Delta \tau (
     \f{J_z}{2}+\f{H}{z} + t+f -v-h), \label{second-flip} \\
   \f{w(a_-,G_{||})}{w(a_+,G_{||})} & = & 1-\Delta \tau (
     \f{2H}{z}-s-e+t+f), \label{third-flip} \\
   \f{w(b,G_\times)}{w(a_+,G_\times)} & = & (\f{|J_x|}{2}-u-g)/s, \label{fourth-flip} \\
   \f{w(b,G_\times)}{w(a_-,G_\times)} & = & (\f{|J_x|}{2}-u-g)/t, \\
   \f{w(a_-,G_\times)}{w(a_+,G_\times)} & = & \f{t}{s}, \\
   \f{w(c,G_=)}{w(b,G_=)} & = & \f{v}{u}, \\
   \f{w(a_-,G_\timesring)}{w(a_+,G_\timesring)} & = & \f{e}{f}. 
   \label{last-flip}
\end{eqnarray}
Given these ratios the flipping probabilities can be gotten from
\be
  P_G(s \rightarrow s\prime) = \min \left[ 
       \f{\prod_p w(s_p\prime,G_p )}{\prod_p w(s_p,G_p)},1 \right].
\ee

\section{Parameter Choices}
There are many possibilities for the choice of parameters, but
not all of them lead to efficient ergodic algorithms. 
To minimize autocorrelation times one must in particular ensure that
the loops generated have a reasonable chance of being flipped. This means that
the ratios in Eqs.~(\ref{first-flip})-(\ref{last-flip}) should be
as close to unity as possible.

Let us first consider $H=0$. The standard loop algorithm is constructed
such that all the ratios Eqs.~(\ref{first-flip})-(\ref{last-flip}) are one,
and by minimizing the weights $w(x,G_\timesring)$:
\begin{eqnarray}
  u_0 & = & \theta (J_z-|J_x|) \f{|J_x|- J_z }{4}  
           +\theta (J_z+|J_x|) \f{|J_x|+ J_z }{4}, \\
  v_0 & = & u_0 , \\
  s_0 & = & \f{|J_x|}{2}-u_0,  \\
  t_0 & = & s_0, \\
  e_0 & = & -\left[1- \theta(J_z+|J_x|) \right] \f{J_z+|J_x|}{2}, \\
  f_0 & = & e_0, \\
  g_0 & = & 0, \\
  h_0 & = & \theta(J_z-|J_x|) \f{J_z-|J_x|}{2}.
\end{eqnarray}
In particular the nonzero parameters for the 
Heisenberg antiferromagnet corresponds to, $u_0=v_0=J/2$, and for the 
XY-model $s_0=t_0=u_0=v_0=J/4$.
For Ising anisotropy, $|J_x|<|J_z|$, certain $G_\timesring$ break-ups must be included as
otherwise some weights will be negative. For extreme anisotropy
$|J_x|=0$ the model is the classical Ising model and the world-lines
are all straight, $s_0=t_0=v_0=0$. In this limit the standard loop algorithm 
above is the Swendsen-Wang algorithm for the Ising model.

Now consider $H \neq 0$. In this case it is not possible to set all 
of the ratios Eqs.~(\ref{first-flip})-(\ref{last-flip}) to unity for any
parameter choices. With the parameter choices for the standard
loop algorithm these ratios are only unity for loops which do not change the 
magnetization when flipped. For loops that can change the magnetization
the total ratio of weights is $\exp(-\beta H \Delta M)$,
where $\Delta M$ is the change in magnetization caused by flipping the loop.
This leads to autocorrelation times that increase exponentially with
$\beta H$. The reason is that the magnetic field
is not taken into account in the loop building process.
The process of changing the magnetization is a competition 
between loosing Zeeman energy and gaining exchange energy. As
the exchange energy is gained in the loop building process, it is inefficient
to build the loops as if the magnetic field was absent.
What happens in the standard loop algorithm is that the number of loops generated which can change 
the magnetization is very small.

An interesting observation is that one can construct an algorithm
where only loops which can change the magnetization are generated.
This choice is
\be
   u=v=e=f=g=h = 0 ,\;  s = t =  \f{|J_x|}{2}, \label{High-H}
\ee
which means that the only break-ups allowed are of the diagonal type.
This algorithm is ergodic, and in the classical Ising limit, $|J_x|=0$,
it corresponds to the standard local Ising model algorithm
in a magnetic field.
One can now ask for which magnetic field this algorithm
minimizes the autocorrelation times. 
Eq. (\ref{first-flip}) is unity for
\be
  \f{H}{z} = \f{|J_x|}{2}+\f{J_z}{2}.
\ee
The important observation is that this value of $H$ is the saturation field, where
almost all plaquettes are of type $a_+$. At this field it is therefore 
not important for the performance of the algorithm that Eqs.~(\ref{second-flip}) and
(\ref{third-flip}) deviate from one.

It is then natural to choose an algorithm
valid for all $H$ which interpolates between the standard algorithm at $H=0$ and the above at the saturation field.
For the Heisenberg antiferromagnet in a magnetic field we thus propose the following algorithm
\begin{eqnarray}
  s=t & = & \f{H}{2z}, \nonumber \\
  u=v & = & \f{J}{2}-\f{H}{2z}. \label{interpolation}
\end{eqnarray} 
For $H > J z$ we use the same algorithm as for $H=J z$. 
Eq.~(\ref{interpolation}) implies that Eqs.~(\ref{first-flip}),(\ref{fourth-flip})-(\ref{last-flip}) becomes unity, whereas Eqs.~(\ref{second-flip})-(\ref{third-flip}) do not.

\section{Numerical Results}
The algorithm was first tested by measuring magnetization of a dimer or two-site $S=1/2$ antiferromagnetic chain. As shown by Kashurnikov {\em et al.}\cite{Worm} the integrated autocorrelation time for the standard loop algorithm increases exponentially with $\beta H$. We have verified this measuring the integrated autocorrelation 
time as described in Ref.1, appendix B. 
For the new algorithm the dimer integrated autocorrelation time is very small ($<2$) down to the lowest temperatures measured (T=.005J) and the magnetization measured agrees excellently with exact results.



\begin{figure}
\begin{center}
\epsfig{file=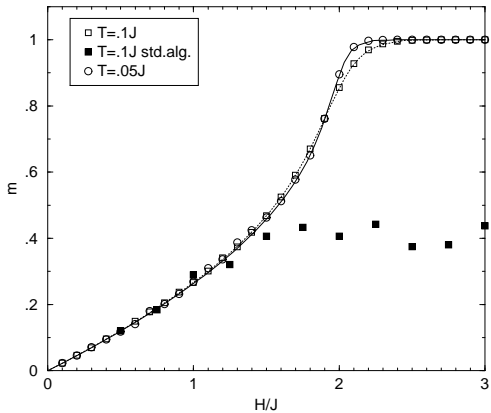,width=6.5cm}
\end{center}
\caption{Magnetization per spin for a spin chain with 64 sites. \label{chain-fig}}
\end{figure}

Fig.~\ref{chain-fig} shows the magnetization per spin of a 64-site antiferromagnetic Heisenberg chain, and illustrates the improvement over the standard Loop algorithm. The results of the modified and standard loop algorithm were obtained using the same number of equilibration and measurement steps ($10^6$), and the lines are exact results obtained using Bethe Ansatz\cite{Takahashi}. 
It is clear that, in contrast to the modified algorithm, results obtained using the standard loop algorithm have not converged for high magnetic fields. 
Close inspection of the data at the lowest temperature reveals that the results of the modified algorithm deviate slightly from the exact results at {\em intermediate} magnetic fields. This deviation which is statistical is caused by increased autocorrelation times which arises because Eqs.~(\ref{second-flip})-(\ref{third-flip}) deviate from unity concomitant with the presence of a significant fraction of $a_{-}$ plaquettes at these fields. For the 64-site chain we measured the integrated autocorrelation time to be a maximum $3\cdot 10^4$ steps at $H/J=1.3$, going down to about 60 steps at low and high fields. It is quite conceivable that a different interpolation scheme than the one chosen in Eq.~(\ref{interpolation}) can reduce these autocorrelation times at intermediate fields.

\begin{figure}
\begin{center}
\epsfig{file=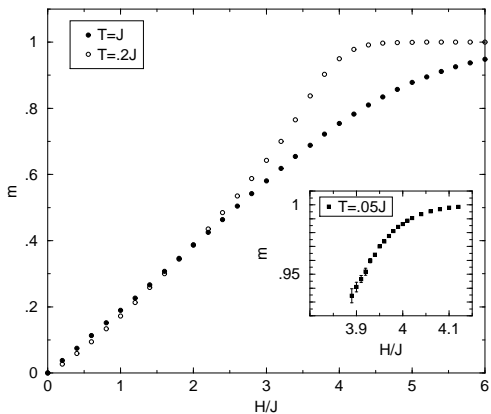,width=6.5cm}
\end{center}
\caption{Magnetization per spin for a plane with $64 \times 64$ sites. \label{plane-fig}}
\end{figure}

Fig.~\ref{plane-fig} shows the full magnetization curve of a 64x64 square lattice antiferromagnet. Typical runs involved $10^7$ steps for equilibration and measurements. The inset shows the high field behavior for very low temperatures on the same lattice. The statistical errors, taking into account the autocorrelation times, are smaller than the symbol size. For the lowest temperature the integrated autocorrelation times reached a maximum of $4 \cdot 10^4$ steps at H/J=2.4 going down to about $4\cdot 10^3$ steps at low and high magnetic fields. 

\begin{figure}
\begin{center}
\epsfig{file=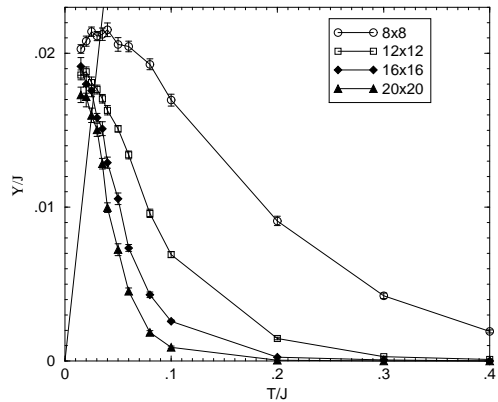,width=6.5cm}
\end{center}
\caption{Helicity modulus as function of temperature for two different system sizes. The line is the Kosterlitz-Thouless-Nelson critical line. \label{helicity-fig}}
\end{figure}

The Heisenberg antiferromagnet in a magnetic field undergoes
a Kosterlitz-Thouless transition at finite temperatures. 
The transition temperature has previously been obtained for weak magnetic fields; $H < .2J$ \cite{Troyer}.  
Fig.~\ref{helicity-fig} shows the
helicity modulus $\Upsilon$, which is the normalized free energy change due to phase twists in the x-y-plane, and which is proportional to the squared spatial winding number of world-lines\cite{Ceperley}, as a function of temperature for four different system sizes at $H=3.95J$. From a finite size analysis\cite{Harada} the transition temperature is found to be $T_c=.020(5) J$. Here a
single-cluster\cite{Evertz} implementation of the algorithm is used. It is expected that more precise estimates for $T_c$ can be obtained using a multi-cluster implementation. 

The author wishes to thank P.A. Lee for useful discussions and 
V. Chudnovsky and U.-J. Wiese for providing source code for the
continuous time loop algorithm.

\end{document}